\documentclass{article}
\usepackage[a4paper, total={6in, 8in}]{geometry}
\usepackage{amsmath}
\usepackage{amsthm}
\usepackage{amssymb}
\usepackage{bbm}
\usepackage{bm}
\usepackage{booktabs}
\usepackage{graphicx,psfrag,epsf}
\usepackage{enumerate}
\usepackage{natbib}
\usepackage{tabularx}
\usepackage{url} 

\newcommand{\blind}{0}

\newtheorem{proposition}{Proposition}

\newcolumntype{Y}{>{\centering\arraybackslash}X}

\begin{document}

\def\spacingset#1{\renewcommand{\baselinestretch}%
{#1}\small\normalsize} \spacingset{1}


\if0\blind
{
  \title{\bf A smooth transition autoregressive model for matrix-variate time series}
  \author{Andrea Bucci\thanks{
    Corresponding author. e-mail: andrea.bucci@unich.it}\hspace{.2cm}\\
    Department of Economics, \\
    Universit\`a degli Studi ``G. d'Annunzio'' Chieti-Pescara, Italy}
  \maketitle
} \fi

\if1\blind
{
  \bigskip
  \bigskip
  \bigskip
  \begin{center}
    {\LARGE\bf Title}
\end{center}
  \medskip
} \fi

\bigskip
\begin{abstract}
In many applications, data are observed as matrices with temporal dependence. Matrix-variate time series modeling is a new branch of econometrics. Although stylized facts in several fields, the existing models do not account for regime switches in the dynamics of matrices that are not abrupt. In this paper, we extend linear matrix-variate autoregressive models by introducing a regime-switching model capable of accounting for smooth changes, the matrix smooth transition autoregressive model. We present the estimation processes with the asymptotic properties demonstrated with simulated and real data.
\end{abstract}

\noindent%
\vfill

\spacingset{1.8} 
\section{Introduction}
Time series processes are present in many fields, including econometrics, finance, biology, ecology, and medicine. The typical classification of time series analysis is between univariate and multivariate frameworks, and both have been extensively studied in the related literature \citep{Hamilton1994, Tsay2014}. There exists a third possible specification in the temporal domain which involves multidimensional datasets \citep{Billio2022}. Of particular interest are bi-dimensional data where the variables are organized in matrices that vary over time. In fact, groups of variables can be observed at different locations and temporal instants, and can be structured in matrix-variate time series \citep{Wang2019}. For instance, a set of financial indicators can be observed for several companies, or a set of environmental variables can be observed at different monitoring stations.

The literature about matrix-variate time series is still scarce. While matrix-valued data has been studied in several papers \cite[see][]{Leng2012, Zhou2013, Zhou2014}, only recently the matrix-variate autoregressive (MAR) model proposed by \cite{Wang2019} and \cite{Chen2021} has opened up new territories for modeling the temporal dependency in matrix-variate problems. Modeling the data as matrices has a twofold advantage with respect to the more used stacked vectorization form: on the one hand, the original matrix structure is preserved and the coefficient matrices can be interpreted row-wisely and column-wisely \citep{Chen2021}; on the other hand, the number of parameters is drastically decreased.

Matrix-variate time series modeling is in its infancy and most of the research in this field has focused on linear autoregressive models. Recently, nonlinear time series models have gained a growing attention, since linear models have several limitations in many empirical applications \citep{Liu2020}. In fact, economic and financial systems have been shown to exhibit both structural and behavioral changes. A natural choice for the modelling of such variables involves the use of a regime-switching mechanism. Over the past 20 years, a rising number of regime-switching models have been proposed to account for nonlinearities, both for univariate and multivariate time series, see \cite{hute13} for a survey. Recently, \cite{Liu2022} have introduced a threshold version of the MAR model that accounts for abrupt changes in the matrices of parameters. In this paper, we suppose that the regime changes can be smooth and we introduce the matrix-variate smooth transition (MSTAR) model inspired by the vector smooth transition autoregressive (VSTAR) model proposed by \cite{teya14}. Although a typical assumption is that the transition variable is weakly stationary, we also analyse the interesting case in which a regime changes coincides with a structural break, meaning that the temporal trend is used as a transition variable. This may help identifying common breaks in matrix-valued problems.

We investigate the estimation process that allows to both estimate the coefficient matrices and the parameters of the transition function. The finite sample performance of the estimators is assessed through a simulation study. We also apply the model to three economic indicators, the Economic Policy Uncertainty, the short- and long-term interest rates, from two countries, the United Kingdom and the United States of America, to both estimate the coefficients in a regime-switching model and assess the presence of a common structural break. Moreover, we apply the model to the daily time series in 2019 of four pollutants from three monitoring stations in the city of London using as a transition variable the average temperature or the precipitation level.

The remainder of the paper is organized as follows. We introduce the general autoregressive model for matrix-variate time series in Section \ref{sec:Model}. Its smooth transition extension is presented in Section \ref{sec:MSTAR}. Section \ref{sec:Estimation} considers the estimation procedures. A simulation study is performed in Section \ref{sec:simulation}, while two real data examples are given in Section \ref{sec:empirical}. Section \ref{sec:Conclusions} concludes.

\section{Matrix autoregressive model}\label{sec:Model}
Consider an $m \times n$ matrix $\mathbf{Y}_t$ observed at time $t$, for $t = 1, \ldots, T$. Let $\text{vec}(\cdot)$ be the vectorization of a matrix by stacking its columns. Assuming a zero-mean process, the traditional vector autoregressive model (VAR) of order 1 is directly applicable for $\text{vec}(\mathbf{Y}_t)$, such that
\begin{equation}\label{eq:vecVAR}
\text{vec}(\mathbf{Y}_t) = \mathbf{\Phi}\text{vec}(\mathbf{Y}_{t-1}) + \text{vec}(\mathbf{E}_t).
\end{equation}
In this model, the role of the lagged variables in column and row is mixed and a single effect cannot be distinguished from the $mn \times mn$ coefficient matrix $\mathbf{\Phi}$, which does no have an assumed structure and is difficult to interpret. To avoid VAR modeling with vectorized data, and to take advantage of the knowledge within the original matrix structure, \cite{Chen2021} propose the matrix autoregressive model (MAR). The first-order version of the model is the following
\begin{equation}\label{eq:MAR}
	\mathbf{Y}_t = \mathbf{A}\mathbf{Y}_{t-1}\mathbf{B}' + \mathbf{E}_t
\end{equation}
where $\mathbf{A}$ and $\mathbf{B}$ are $m \times m$ and $n \times n$ autoregressive coefficient matrices, and $\mathbf{E}_t$ is a $m \times n$ matrix white noise such that $E(\mathbf{E}_t) = 0$, $E(\mathbf{E}_t \otimes \mathbf{E}_t') = \mathbf{\Sigma}_t$, and $E(\mathbf{E}_t \otimes \mathbf{E}_{t'}') = 0$, for $t \neq t'$ \citep{Samadi2014}, where $\mathbf{\Sigma}_t$ is a $mn \times mn$ covariance matrix. 
It is further assumed that $\mathbf{E}_t$ has a zero-mean matrix normal distribution \citep{Guptar1999}. Although throughout the paper we focus on a single-lag model, this can be extended to a $p$ order model.

\begin{proposition}
	(Weak Stationarity) A stochastic matrix process ($\mathbf{Y}_t$) is (weakly) stationary if it possesses finite first and second moments which are constant through time. A $m \times n$ time series matrix, $\mathbf{Y}_t$, is stationary if, $\forall t \in [0,T]$ and $\forall h \in \mathbbm{N}$ (the natural numbers), it satisfies
	\begin{align}
	&E[\mathbf{Y}_t]  = \mathbf{M} < \infty \label{eq:Prop1a}\\
	&E[\mathbf{Y}_{t+h}\otimes \mathbf{Y}_t'] = \mathbf{\Sigma}(h) < \infty, \label{eq:Prop1b}
	\end{align}
where $\mathbf{M}$ is a mean $m \times n$ matrix, and $\mathbf{\Sigma}(h)$ is the lag function.
\end{proposition}

Therefore, the assumption of a weakly stationarity (or second-order stationarity) implies that the autocovariances of the matrix process are finite and only depend on the lag $h$ but not on time $t$. It should be noted that, differently from the univariate case, we do not have that $\mathbf{\Sigma}(h) = \mathbf{\Sigma}(-h)$. 

The MAR(1) model in \eqref{eq:MAR} can be represented in the form of a vector autoregressive model
\begin{equation}\label{eq:vecMAR}
\text{vec}(\mathbf{Y}_t) = (\mathbf{B}\otimes \mathbf{A})\text{vec}(\mathbf{Y}_{t-1}) + \text{vec}(\mathbf{E}_t)
\end{equation}
where $\otimes$ denotes the Kronecker product. The representation \eqref{eq:vecMAR} means that the MAR(1) model can be viewed as a special case of the classical VAR(1) model in \eqref{eq:vecVAR}, with its autoregressive coefficient matrix given by a Kronecker product. On the other hand, comparing \eqref{eq:vecVAR} and \eqref{eq:vecMAR}, we see that the MAR(1) requires $m^2 + n^2$ coefficients as the entries of $\mathbf{A}$ and $\mathbf{B}$, while an unrestricted VAR(1) needs $m^2n^2$ coefficients for $\mathbf{\Phi}$. It follows that the number of parameters may explode when both $m$ and $n$ are large.

The MAR(1) model has an identifiability issue that concerns the coefficient matrices $\mathbf{A}$ and $\mathbf{B}$. In fact, the model remains unchanged if the two matrices are divided and multiplied by the same nonzero constant respectively. To avoid ambiguity, \cite{Chen2021} normalize $\mathbf{A}$ so that its Frobenius norm is one, while \cite{Wang2019} and \cite{Liu2022} use a dynamic factor model. On the other hand, the uniqueness always holds for the Kronecker product $\mathbf{B} \otimes \mathbf{A}$. 

The error matrix sequence, $\mathbf{E}_t$, is assumed to be a matrix white noise, \textit{i.e.}, there is no correlation between $\mathbf{E}_t$ and $\mathbf{E}_s$ as long as $t \neq s$. But $\mathbf{E}_t$ is still allowed to have concurrent correlations among its own entries. As a matrix, its covariance form a 4-dimensional tensor, which is difficult to express.

\section{Matrix smooth transition autoregressive model}\label{sec:MSTAR}
One way to introduce some nonlinearity in Eq. \eqref{eq:vecVAR} is generalising to the multivariate framework the threshold autoregressive model proposed by \cite{quan58}, as done in \cite{Tsay1998}. The extension in the matrix-variate framework is straightforward and is presented in \cite{Liu2022} in combination with a dimensionality reduction method through factors.

Another possible nonlinear extension of model \eqref{eq:vecVAR} is the vector smooth transition regression (VSTR), firstly appeared in \cite{andva98} and then formalized in \cite{teya14}. Supposing the absence of exogenous variables and a single lag, the 2-regime vector smooth transition autoregressive (VSTAR) model can be defined as follows
\begin{equation}\label{eq:VSTR}
	\text{vec}(\mathbf{Y}_t) = \mathbf{\Phi}_0\text{vec}(\mathbf{Y}_{t-1}) + \mathbf{G}(\mathbf{\gamma}, \mathbf{c}; \mathbf{s}_t)\left(\mathbf{\Phi}_1\text{vec}(\mathbf{Y}_{t-1})\right) + \text{vec}(\mathbf{E}_t)
\end{equation}
where $\mathbf{\Phi}_{k}$ are $mn \times mn$ parameter matrices, for $k = 0,1$, and $\mathbf{G}(\mathbf{\gamma}, \mathbf{c}; \mathbf{s}_t)$ is an $mn \times mn$ diagonal matrix, such that
\begin{equation}\label{eq:Gt}
\mathbf{G}(\mathbf{\gamma}, \mathbf{c}; \mathbf{s}_t) = \text{diag}\left\{g_1(\gamma_1, c_1; s_{1,t}), \ldots, g_{mn}(\gamma_{mn}, c_{mn}; s_{mn, t})\right\},
\end{equation}
where $g(\cdot)$ is typically a standard logistic function, $g_i(\gamma_i, c_i; s_{i,t}) =\left[1 + \exp\left\{-\gamma_{i}\left(s_{i,t}-c_{i}\right)\right\}\right]^{-1}$, and $s_{i,t}$, for $i = 1, \ldots, mn$, is a weakly stationary transition variable usually chosen among the set of variables in $\text{vec}(\mathbf{Y}_{t-1})$, although exogenous variables are possible as well \citep{He2008}. Special cases of the model in \eqref{eq:VSTR} foresee a common transition variable for all the $mn$ equations or, more strictly, a unique transition function, such that
\begin{equation}\label{eq:VSTRspecial}
	\text{vec}(\mathbf{Y}_t) = \mathbf{\Phi}_0\text{vec}(\mathbf{Y}_{t-1}) + g(\gamma, c; s_t)\left(\mathbf{\mu}_1 + \mathbf{\Phi}_1\text{vec}(\mathbf{Y}_{t-1})\right) + \text{vec}(\mathbf{E}_t),
\end{equation}
where a single function drives the regime changes of the autoregressive parameters. Extending to the matrix-variate framework the smooth transition version the model in \eqref{eq:MAR} may imply introducing a regime-switching either for the parameters related to the column-wise effect, those related to the row-wise effect or both. In this case, we could have several specifications. 

The regime change may involve only the row-wise interactions, matrix $\mathbf{C}$ in the following equation, with a transition function for each element in row, such that model \eqref{eq:MAR} becomes
\begin{equation*}
	\mathbf{Y}_t = \mathbf{A}\mathbf{Y}_{t-1}\mathbf{B}_0' + \mathbf{G}_{t,C}(\mathbf{\gamma}_C, \mathbf{c}_C; \mathbf{s_t}_C)\mathbf{C}\mathbf{Y}_{t-1} \mathbf{D}' + \mathbf{E}_t,
\end{equation*}
where $\mathbf{G}_{t,C}$ is an $m \times m$ diagonal matrix with standard logistic function on its diagonal, and the transition variables can be either different or common among the $m$ transition functions.

Alternatively, it can be assumed that the transition affects only the column-wise interactions. This means that a diagonal transition matrix, $\mathbf{G}_{t,D}$, enters in the model as follows
\begin{equation*}
	\mathbf{Y}_t = \mathbf{A}\mathbf{Y}_{t-1}\mathbf{B}_0' + \mathbf{C}\mathbf{Y}_{t-1} \mathbf{D}'\mathbf{G}_{t,D}(\mathbf{\gamma}_D, \mathbf{c}_D; \mathbf{s_t}_D) + \mathbf{E}_t.
\end{equation*}

By supposing an independent regime-switching function for each of the two parameters matrices, the more general model is
\begin{equation*}
	\mathbf{Y}_t = \mathbf{A}\mathbf{Y}_{t-1}\mathbf{B}_0' + \mathbf{G}_{t,C}(\mathbf{\gamma}_C, \mathbf{c}_C; \mathbf{s_t}_C)\mathbf{C}\mathbf{Y}_{t-1} \mathbf{D}'\mathbf{G}_{t,D}(\mathbf{\gamma}_D, \mathbf{c}_D; \mathbf{s_t}_D) + \mathbf{E}_t.
\end{equation*}
However, the interpretation of the parameters in such a model becomes extremely difficult, since the transition variables and the parameters of the function can differ for each element of $\mathbf{C}$ and $\mathbf{D}$. A simpler version of the aforementioned models foresees that the transition variables, as well as the parameters of the transition functions, are the same for all the equations \citep{hute13}. For its simplicity, its popularity in most of the macroeconomic problems and its similarity with a threshold autoregressive model as the one introduced in \cite{Liu2022}, we use this specification to extend in a matrix-variate context model \eqref{eq:VSTRspecial}. Supposing a two-regime model (\textit{i.e.}, a single transition function), we extend model \eqref{eq:MAR} by introducing an additive nonlinear component, such that
\begin{equation}\label{eq:SMAR}
	\mathbf{Y}_t = \mathbf{A}\mathbf{Y}_{t-1}\mathbf{B}' + g_t(\gamma, c; s_t) (\mathbf{C}\mathbf{Y}_{t-1} \mathbf{D}') + \mathbf{E}_t,
\end{equation}
where $g_t(\gamma, c,; s_t)$ is, as before, a standard logistic function (for ease of notation we will refer to this function as $g_t$). Obviously, more than two regimes are also allowed even if in practice a number of regimes greater than 3 is of scarce interest. This matrix smooth transition autoregressive (MSTAR) model implies that the conditional mean of the entire matrix changes the regime based on the values of the unique transition variable, $s_t$. This is not entirely different from what happens in practice. For example, let think to the financial variables cited by \cite{Wang2019} for matrix-variate problems. In the case of a shock on the financial markets, all the variables changes their behaviour, especially in terms of persistence \citep{Angelini2019}. Nevertheless, finding a common transition variable that drives the dynamics of the entire matrix is not always easy, especially when the entries of the matrix are different indices. For this reason, a more interesting candidate transition variable is the normalized temporal trend, $s_t = t/T$. In this case, model \eqref{eq:VSTR} becomes a linear vector regression with a smooth structural break \citep{He2008}. It follows that, if we use $s_t = t/T$ as a transition variable in the MSTAR model proposed in this article, the estimation of the threshold parameter $c$ implies a single structural break detection for all the matrix-valued time series.

As before, the model in Eq. \eqref{eq:SMAR} can be specified in a vector form, such that
\begin{equation}\label{eq:vecSMAR}
	\text{vec}(\mathbf{Y}_t) = (\mathbf{B}\otimes\mathbf{A})\text{vec}(\mathbf{Y}_{t-1}) + g_t ((\mathbf{D}\otimes \mathbf{C}) \text{vec}(\mathbf{Y}_{t-1})) + \text{vec}(\mathbf{E}_t).
\end{equation}

This permits to define the conditions on the coefficient matrices to guarantee the stationarity of the process and the consistency of the estimates. Assuming that $\rho(\cdot)$ is the spectral radius operator, the model in Eq. \eqref{eq:vecSMAR} admits solutions only if the spectral radius of $\mathbf{B} \otimes \mathbf{A}$ and $\mathbf{D} \otimes \mathbf{C}$ is strictly minor than 1. Therefore, we have that:
\begin{proposition}
The MSTAR(1) model in Eq. \eqref{eq:SMAR} is stationary if $\rho(\mathbf{A}) \cdot \rho(\mathbf{B}) <1$ and $\rho(\mathbf{C}) \cdot \rho(\mathbf{D}) <1$.
\end{proposition}
This Proposition follows from the condition on the roots in the polynomial $\text{det}\left(\mathbf{I} - (\mathbf{B}\otimes \mathbf{A}) \text{vec}(\mathbf{Y}_{t-1})\right)$ which should lie outside the unitary circle in model \eqref{eq:vecMAR}.

\subsection{Estimation of the MSTAR model}\label{sec:Estimation}
In a general MAR model as the one in Eq. \eqref{eq:MAR}, assuming that the entries of $\mathbf{E}_t$ are i.i.d. normal with mean zero and a constant variance, the estimates of the parameters in Eq. \eqref{eq:MAR} are the solution of the least squares problem
\begin{equation}
	\underset{\mathbf{A}, \mathbf{B}}{\min} \sum_t \parallel\mathbf{Y}_t - \mathbf{A}\mathbf{Y}_{t-1}\mathbf{B}'\parallel_F^2.
\end{equation}
\cite{Chen2021} show that the estimates of the matrices of parameters can be obtained either through a projection method, iterative least squares and maximum likelihood. In the estimation process of the nonlinear extension of the MAR, we focus on the latter method.

The estimation of the MSTAR model is more complicated than a simple MAR model for the presence of the threshold and the slope parameters. Moreover, the splitting algorithm used for the iterative least squares estimates used for a matrix-variate threshold autoregressive model \citep{Liu2022} is not applicable. One way to solve these shortcomings is using the same algorithm proposed in \cite{hute13} which foresees the minimization of the objective function conditionally on $\gamma$ and $c$. Let the optimization problem be
\begin{equation}\label{eq:minMSTAR}
	\underset{\mathbf{A}, \mathbf{B}, \mathbf{C}, \mathbf{D}}{\min} \sum_t \parallel\mathbf{Y}_t - \mathbf{A}\mathbf{Y}_{t-1}\mathbf{B}' - g_t\mathbf{C} \mathbf{Y}_{t-1}\mathbf{D}'\parallel_F^2,
\end{equation}
which leads to the following loss function to be minimized:
\begin{align}\label{eq:minMSTAR2}
	Q &= \sum_t \left(\mathbf{Y}_t - \mathbf{A}\mathbf{Y}_{t-1}\mathbf{B}' - g_t \mathbf{C}\mathbf{Y}_{t-1}\mathbf{D}'\right)'\left(\mathbf{Y}_t - \mathbf{A}\mathbf{Y}_{t-1}\mathbf{B}' - g_t \mathbf{C}\mathbf{Y}_{t-1}\mathbf{D}'\right)\\
	&= \sum_t (\mathbf{Y}_t'\mathbf{Y}_t - 2 \mathbf{Y}_t'\mathbf{A}\mathbf{Y}_{t-1}\mathbf{B}' - 2g_t\mathbf{Y}_t'\mathbf{C}\mathbf{Y}_{t-1}\mathbf{D}' + 2g_t\mathbf{B}\mathbf{Y}_{t-1}'\mathbf{A}'\mathbf{C}\mathbf{Y}_{t-1}\mathbf{D}' \nonumber\\
	&+ \mathbf{B}\mathbf{Y}_{t-1}'\mathbf{A}'\mathbf{A}\mathbf{Y}_{t-1}\mathbf{B}' + g_t^2\mathbf{D}\mathbf{Y}_{t-1}'\mathbf{C}'\mathbf{C}\mathbf{Y}_{t-1}\mathbf{D}'). \nonumber
\end{align}
It is clear that if all the coefficients are zeros, than the parameters $\gamma$ and $c$ are non-identifiable. A typical assumption to solve this issue is that not all the coefficients are zeros. Therefore, conditionally on the values of the transition function, the minimization of the loss function in Eq. \eqref{eq:minMSTAR2} is guaranteed to have at least one global minimum and can be simply solved by taking its first derivatives where $g_t$ is treated as a known scalar. As in the case of the MAR, the solution of \eqref{eq:minMSTAR} is found by iteratively updating one at a time the matrices $\hat{\mathbf{A}}$, $\hat{\mathbf{B}}$, $\hat{\mathbf{C}}$, and $\hat{\mathbf{D}}$, while keeping the other matrices fixed, starting with some initial matrices. Using the first derivatives, the iteration of updating one matrix given the others is
\begin{align*}
&\mathbf{A} \leftarrow \left(\sum_t\mathbf{Y}_t\mathbf{B}\mathbf{Y}_{t-1}' - g_t\mathbf{C}\mathbf{Y}_{t-1}\mathbf{D}'\mathbf{B}\mathbf{Y}_{t-1}'\right)\left(\sum_t\mathbf{Y}_{t-1}\mathbf{B}'\mathbf{B}\mathbf{Y}_{t-1}'\right)^{-1}\\
&\mathbf{B} \leftarrow \left(\sum_t\mathbf{Y}_t'\mathbf{A}\mathbf{Y}_{t-1} - g_t\mathbf{D}\mathbf{Y}_{t-1}'\mathbf{C}'\mathbf{A}\mathbf{Y}_{t-1}\right)\left(\sum_t \mathbf{Y}_{t-1}'\mathbf{A}'\mathbf{A}\mathbf{Y}_{t-1}\right)^{-1}\\
&\mathbf{C} \leftarrow \left(\sum_tg_t \mathbf{Y}_t\mathbf{D}\mathbf{Y}_{t-1}' - g_t\mathbf{A}\mathbf{Y}_{t-1}\mathbf{B}'\mathbf{D}\mathbf{Y}_{t-1}'\right)\left(\sum_t g_t^2\mathbf{Y}_{t-1}\mathbf{D}'\mathbf{D}\mathbf{Y}_{t-1}'\right)^{-1}\\
&\mathbf{D} \leftarrow \left(\sum_t g_t\mathbf{Y}_t'\mathbf{C}\mathbf{Y}_{t-1} - g_t \mathbf{B}\mathbf{Y}_{t-1}'\mathbf{A}'\mathbf{C}\mathbf{Y}_{t-1}\right) \left(\sum_t g_t^2 \mathbf{Y}_{t-1}'\mathbf{C}'\mathbf{C}\mathbf{Y}_{t-1}\right)^{-1}.
\end{align*}
Once obtained the iterative least squares estimates of the autoregressive coefficient matrices, the optimization problem foresees the minimization of the loss in Eq. \eqref{eq:minMSTAR} with respect to $\gamma$ and $c$. Finding the optimum of such a function is performed analytically. In practice, the numerical algorithm may be slow and may converge to some local minimum. For this reason, we apply the same method proposed by \cite{hute13} which implies the use of a grid search among a set of possible values for $\gamma$ and $c$. We construct a discrete grid in the parameter space of the transition function parameters and we estimate as before the autoregressive coefficients, conditionally on the pair of values on the grid. The estimation of the parameters in $\mathbf{A}$, $\mathbf{B}$, $\mathbf{C}$ and $\mathbf{D}$ is carried out for all the values in the grid and the residuals sum of squares computed as follows
\begin{equation}
\text{SSQ} = \text{tr}\left(\sum_t \text{vec}(\mathbf{Y}_t - \mathbf{A}\mathbf{Y}_{t-1}\mathbf{B}' - g_t\mathbf{C}\mathbf{Y}_{t-1}\mathbf{D}')'\text{vec}(\mathbf{Y}_t - \mathbf{A}\mathbf{Y}_{t-1}\mathbf{B}' - g_t\mathbf{C}\mathbf{Y}_{t-1}\mathbf{D}')\right)
\end{equation}
is collected. We then choose as starting values for the iterative least squares presented above the pair of values $(\gamma, c)$ that produces the smallest $\text{SSQ}$. Once obtained the estimates of the autoregressive coefficient parameters, the algorithm proceeds with the estimation of $\gamma$ and $c$ fixing all the parameters in $\mathbf{A}$, $\mathbf{B}$, $\mathbf{C}$ and $\mathbf{D}$. This continues until convergence.

\section{Simulation study}\label{sec:simulation}

To study the finite-sample properties of the model introduced in this article, we perform a simulation study in which we generate 100 replications using different settings in terms of dimensionality of the matrix (\textit{i.e.}, with different $m$ and $n$), the length of the time series, with $T = 400, 600,1000$, and the slope parameter, $\gamma = 10,20$.

The simulated $\mathbf{Y}_t$ is generated from model \eqref{eq:SMAR}, where $\mathbf{A}$ and $\mathbf{B}$ are diagonal matrices with 0.20 entries on their diagonal, while $\mathbf{C}$ and $\mathbf{D}$ are computed as a diagonal matrix with entries 0.75. For all the replications we use the same coefficient matrices and we let $\mathbf{E}_t$ vary. This guarantees that $\rho(\mathbf{A})\cdot \rho(\mathbf{B}) <1$ and $\rho(\mathbf{C})\cdot \rho(\mathbf{D}) <1$, then all the DGPs are stationary. Since the iterative algorithm for the estimation of the matrices foresees a set of initial values for the matrices $\mathbf{B}$, $\mathbf{C}$ and $\mathbf{D}$, we sample the value of $\mathbf{B}$ from a uniform distribution $\mathcal{U}(0, 0.2)$, while the entries of the initial $\mathbf{C}$ and $\mathbf{D}$ are sampled from a uniform distribution $\mathcal{U}(0, 0.5)$. In addition, we let $\gamma$ range between 1 and 50 and $c$ between 0 and 1 to compute the searching grid algorithm.

The innovation matrix, $\mathbf{E}_t$, is sampled from a matrix normal distribution, with the $mn \times mn$ covariance matrix equal to $\mathbf{\Sigma} = \mathbf{I}_m\otimes \mathbf{I}_n$. We choose to use as a transition variable the scaled temporal trend, \textit{i.e.}, $s_t = t/T$, and we set $c = 0.65$, while different values of $\gamma$ are used in different simulation settings, to understand whether the abruptness of changes influences the estimation of the transition parameters.

We also estimate the stacked version of the model, this means estimating a VLSTAR model on $\text{vec}(\mathbf{Y}_t)$, and for both the models we compute the box-plot among the 100 replications of the following loss functions
\begin{equation}\label{eq:measure1}
	\parallel\hat{\mathbf{B}} \otimes \hat{\mathbf{A}} - \mathbf{B} \otimes \mathbf{A}\parallel_F^2
\end{equation}
and
\begin{equation}\label{eq:measure2}
	\parallel\hat{\mathbf{D}} \otimes \hat{\mathbf{C}} - \mathbf{D} \otimes \mathbf{C}\parallel_F^2.
\end{equation}

Finally, we report in Table \ref{table:MSE} the mean squared errors (MSE) for the threshold and the slope parameters calculated for each simulation setting and computed as $R^{-1}\sum_{r=1}^{R = 100}\left(\hat{\theta}_r - \theta_0\right)^2$, where $\hat{\theta}_r$ is the estimate of the parameter, either $c$ or $\gamma$, from the $r$-th replicate and $\theta_0$ is the true parameter.

\begin{figure}[h]
	\centering
	\caption{Comparison of the estimators for the MSTAR and VLSTAR models. The rows identify an increasing matrix of $(m,n) = (2,3), (4,6)$, while the number of observations in the sample grows by column. For each subfigure, the plot on the left denotes the first regime, while the plot on the right reports the measures for the second regime.} 
	\label{fig:Coefficients}
	\begin{tabular}{ccc}
\multicolumn{3}{c}{$\gamma = 10$}\\
\includegraphics[scale=0.25]{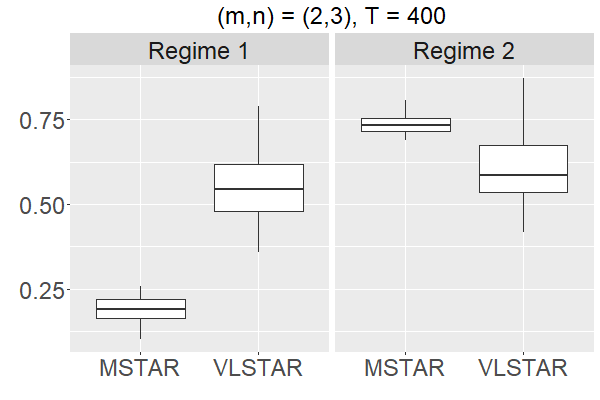} &\includegraphics[scale=0.25]{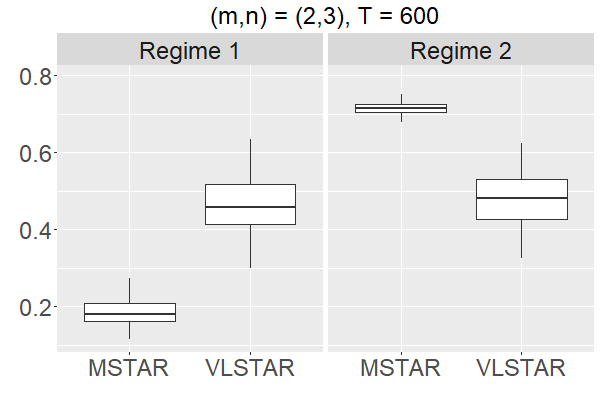}&\includegraphics[scale=0.25]{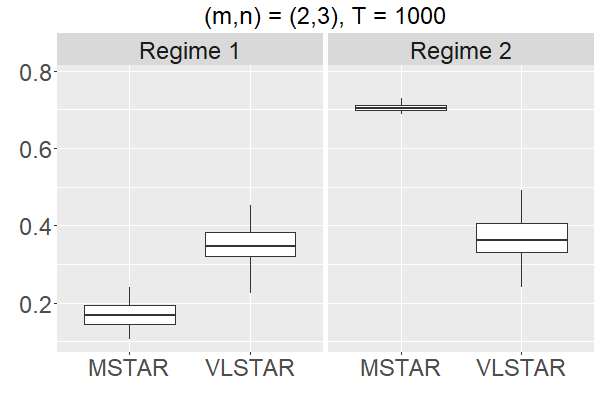}\\
\includegraphics[scale=0.25]{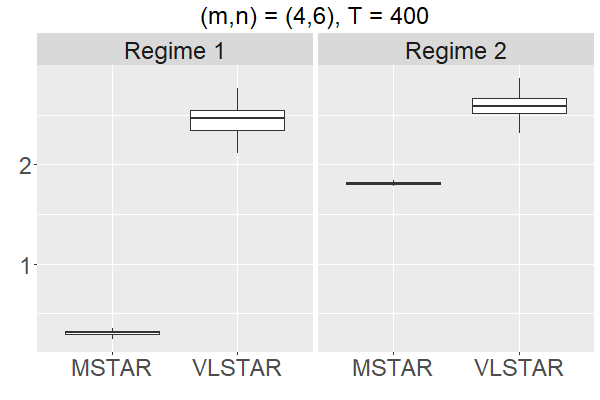} &\includegraphics[scale=0.25]{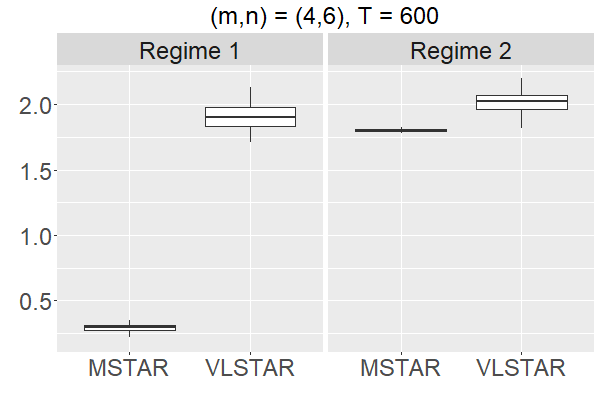} &\includegraphics[scale=0.25]{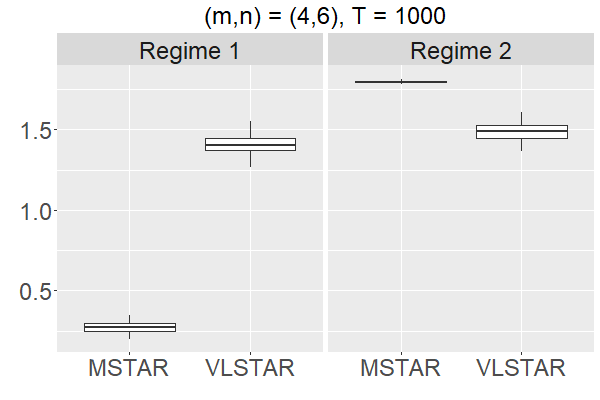}\\
\multicolumn{3}{c}{$\gamma = 20$}\\
\includegraphics[scale=0.25]{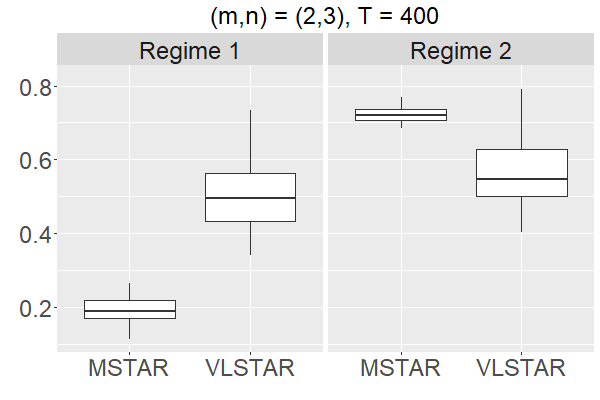} &\includegraphics[scale=0.25]{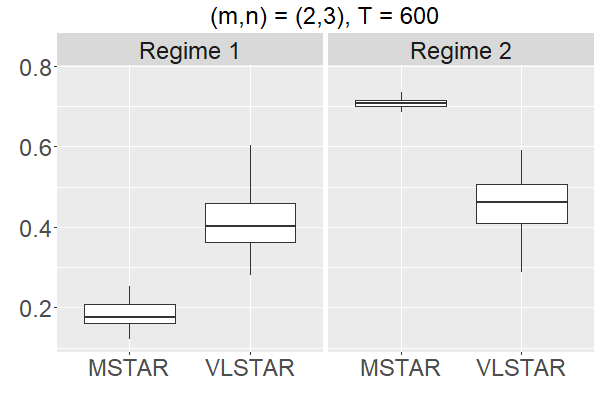}&\includegraphics[scale=0.25]{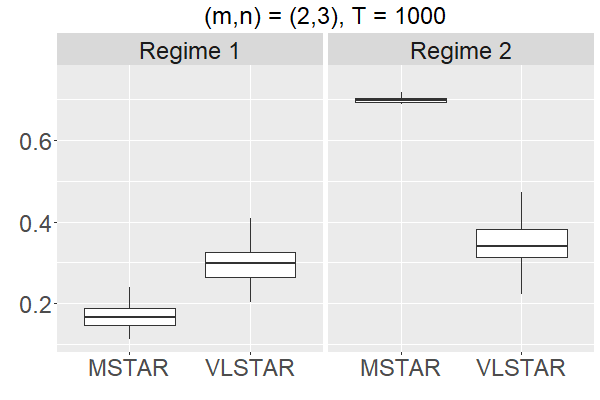}\\
\includegraphics[scale=0.25]{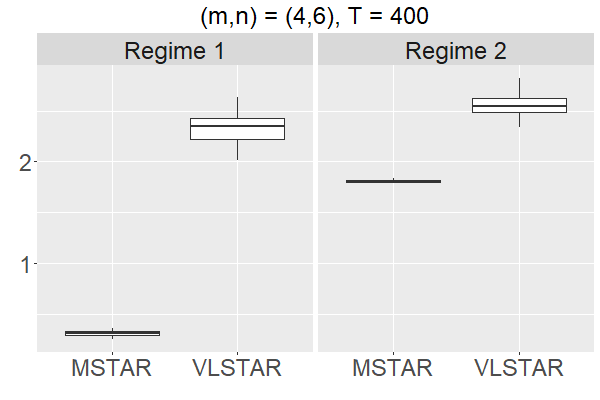} &\includegraphics[scale=0.25]{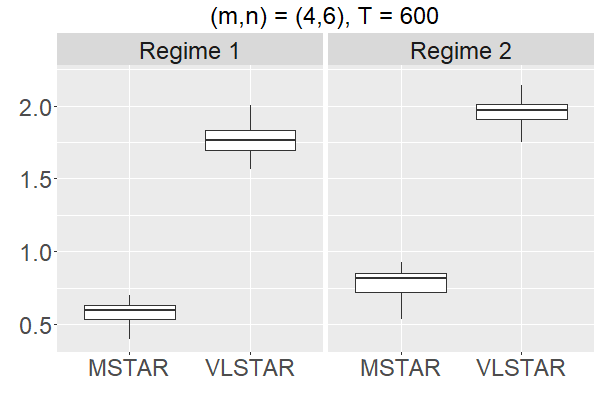} &\includegraphics[scale=0.25]{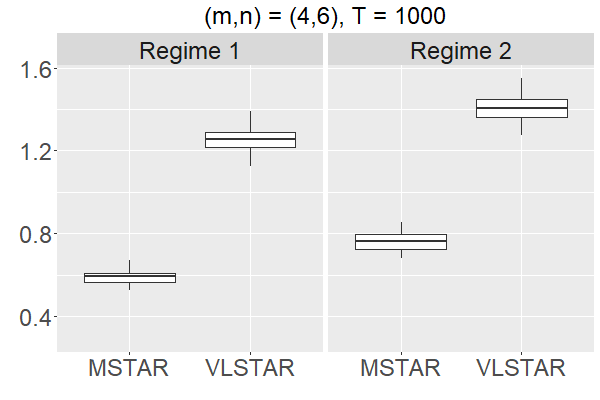}\\
	\end{tabular}
\end{figure}

\begin{figure}[h]
	\centering
	\caption{Comparison of the differences between the estimated threshold parameter and the real parameter ($c = 0.65$) for the MSTAR and VLSTAR models. The rows identify an increasing matrix of $(m,n) = (2,3), (4,6)$, while the number of observations in the sample grows by column. For each subfigure, the plot on the left denotes the first regime, while the plot on the right reports the measures for the second regime.} 
	\label{fig:Thresholds}
	\begin{tabular}{ccc}
		\multicolumn{3}{c}{$\gamma = 10$}\\
		\includegraphics[scale=0.25]{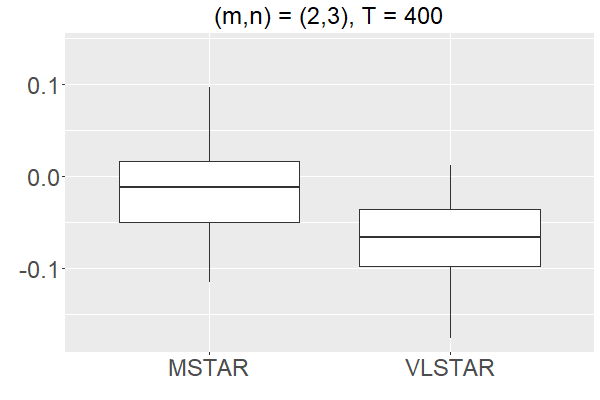} &\includegraphics[scale=0.25]{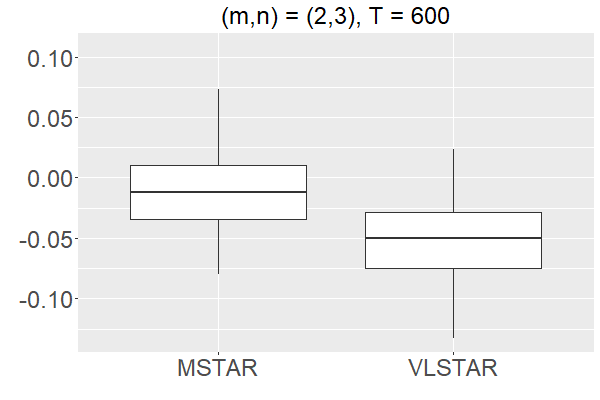} &\includegraphics[scale=0.25]{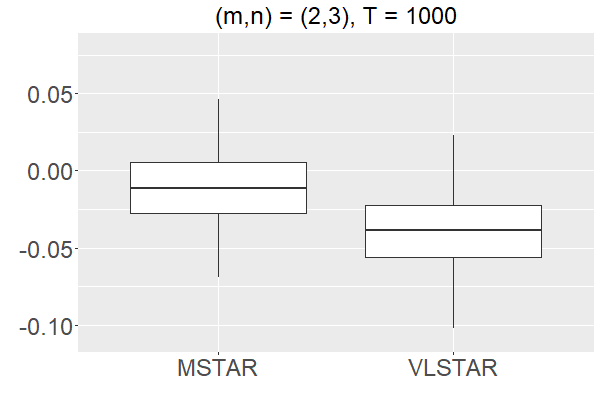}\\
		\includegraphics[scale=0.25]{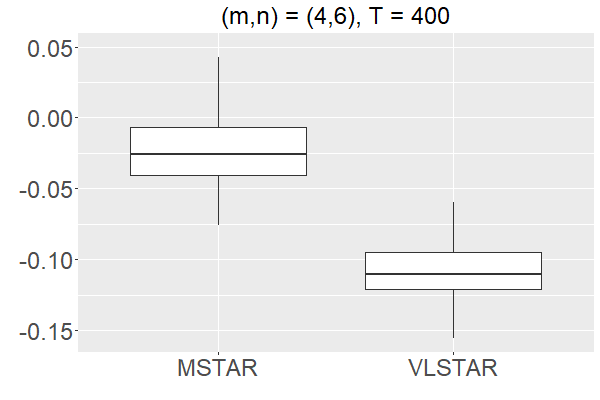} &\includegraphics[scale=0.25]{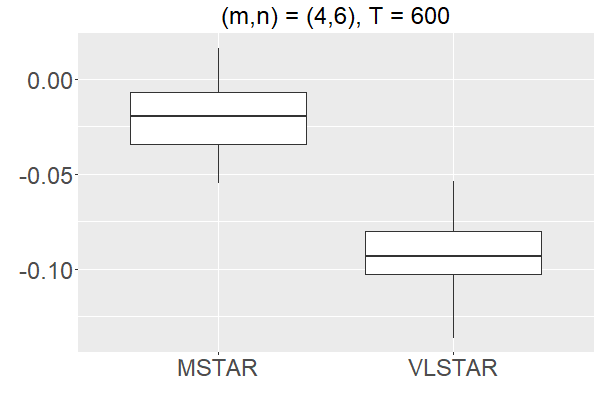} &\includegraphics[scale=0.25]{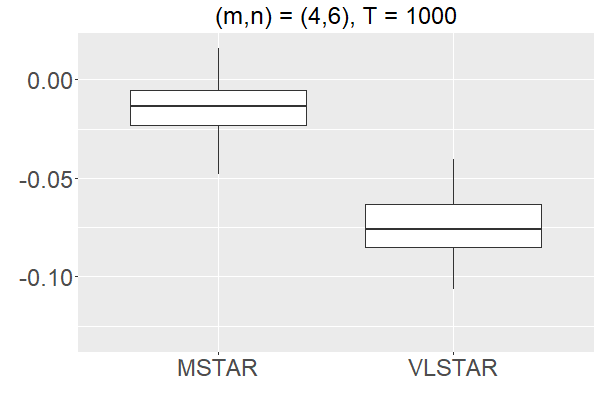}\\
		\multicolumn{3}{c}{$\gamma = 20$}\\
		\includegraphics[scale=0.25]{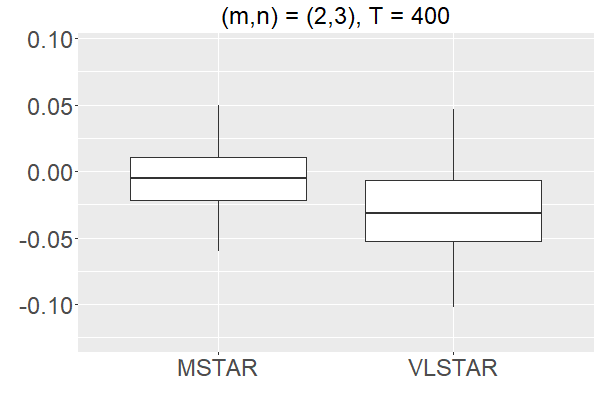} &\includegraphics[scale=0.25]{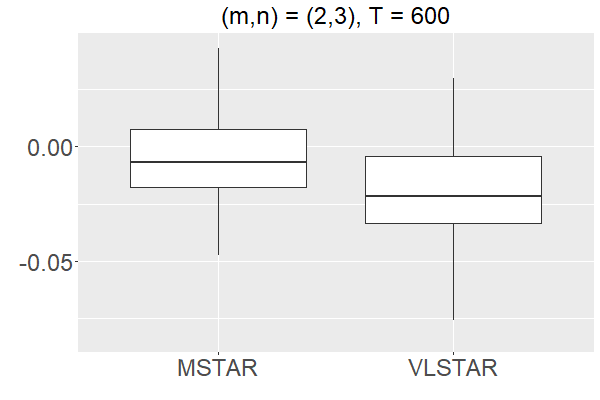} &\includegraphics[scale=0.25]{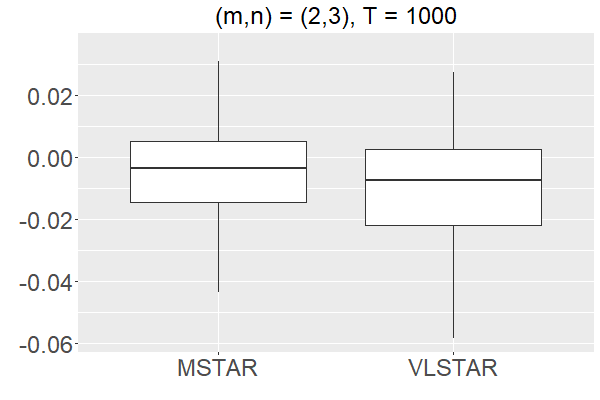}\\
		\includegraphics[scale=0.25]{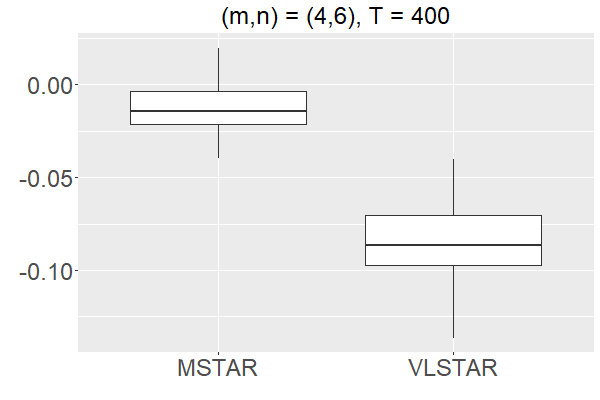} &\includegraphics[scale=0.25]{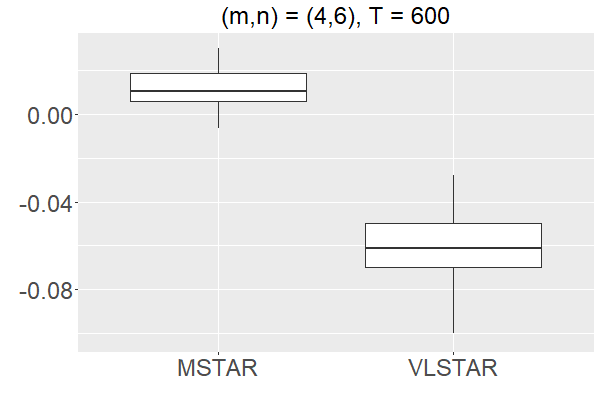} &\includegraphics[scale=0.25]{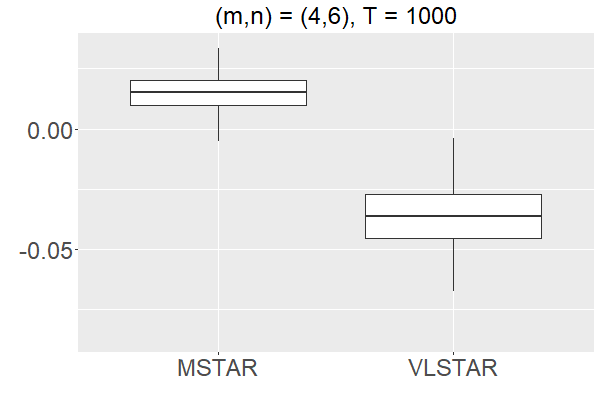}\\
	\end{tabular}
\end{figure}

\begin{figure}[h]
	\centering
	\caption{Comparison of the differences between the estimated slope parameter and the real parameter ($\gamma = 10,20$) for the MSTAR and VLSTAR models. The rows identify an increasing matrix of $(m,n) = (2,3), (4,6)$, while the number of observations in the sample grows by column. For each subfigure, the plot on the left denotes the first regime, while the plot on the right reports the measures for the second regime.} 
	\label{fig:Gammas}
	\begin{tabular}{ccc}
		\multicolumn{3}{c}{$\gamma = 10$}\\
		\includegraphics[scale=0.25]{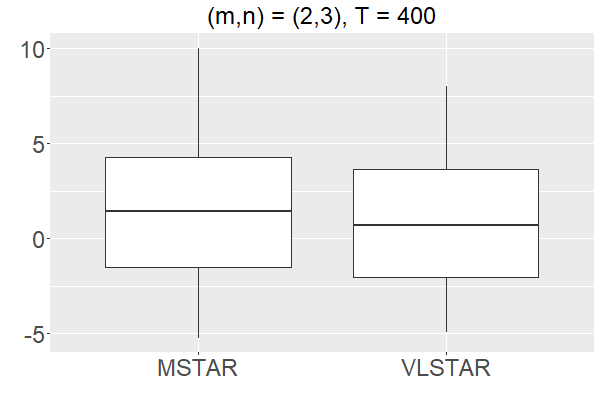} &\includegraphics[scale=0.25]{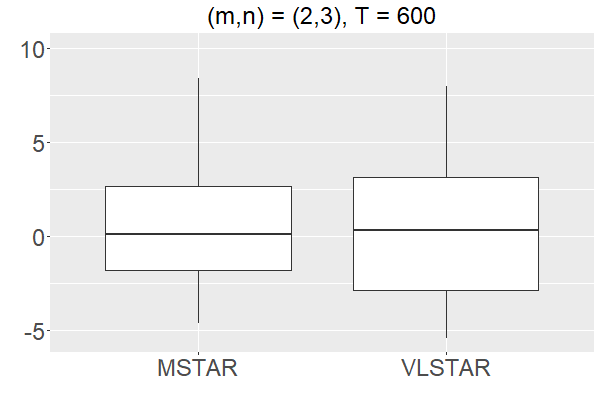} &\includegraphics[scale=0.25]{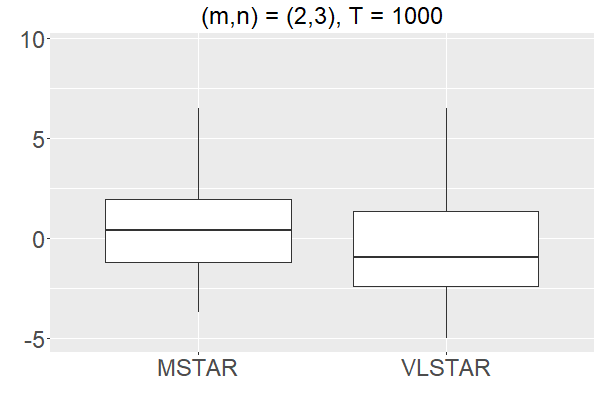}\\
		\includegraphics[scale=0.25]{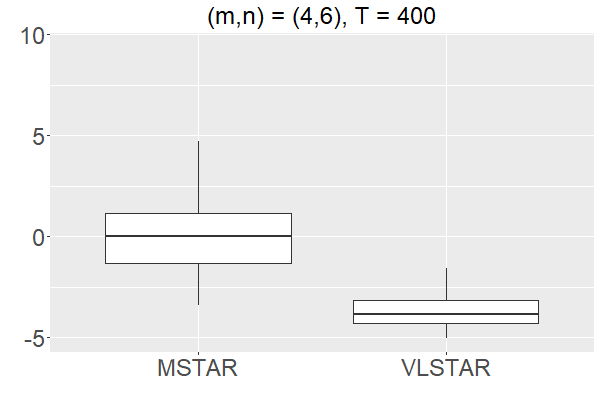} &\includegraphics[scale=0.25]{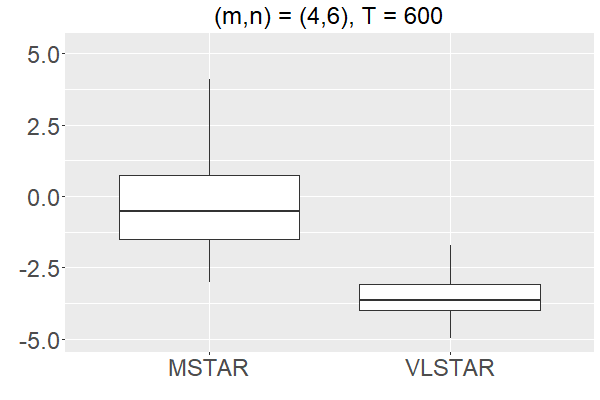} &\includegraphics[scale=0.25]{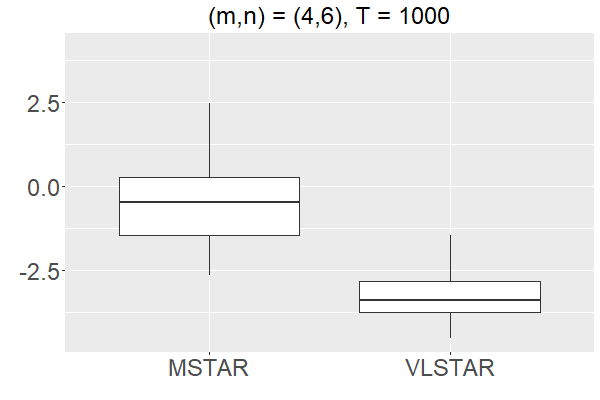}\\
		\multicolumn{3}{c}{$\gamma = 20$}\\
		\includegraphics[scale=0.25]{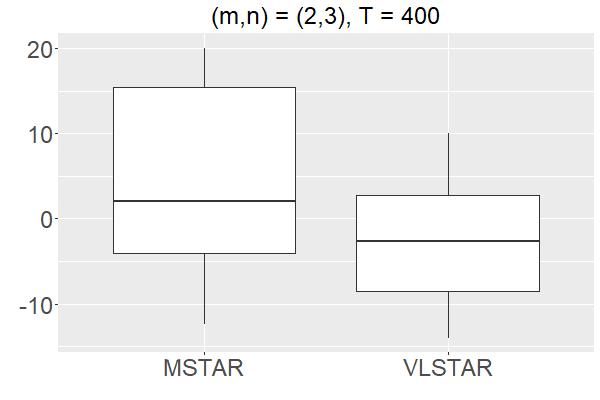} &\includegraphics[scale=0.25]{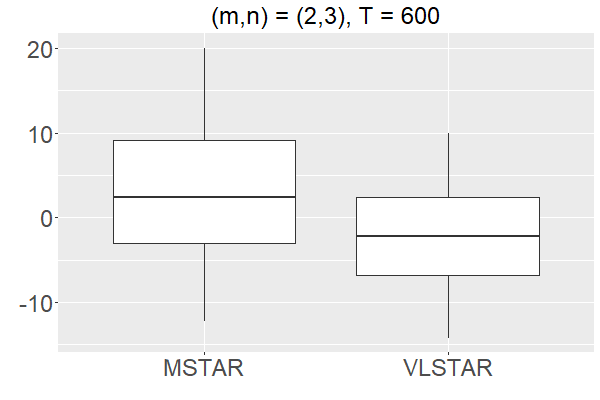} &\includegraphics[scale=0.25]{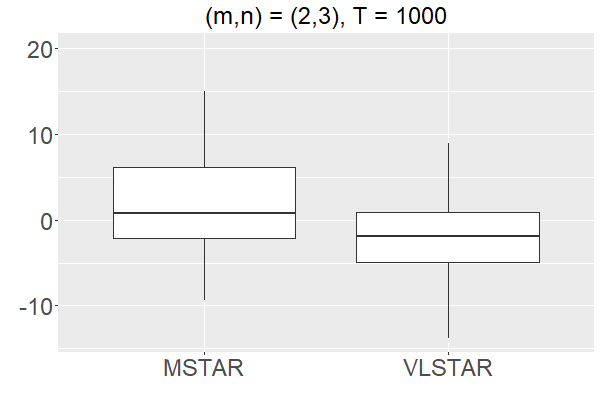}\\
		\includegraphics[scale=0.25]{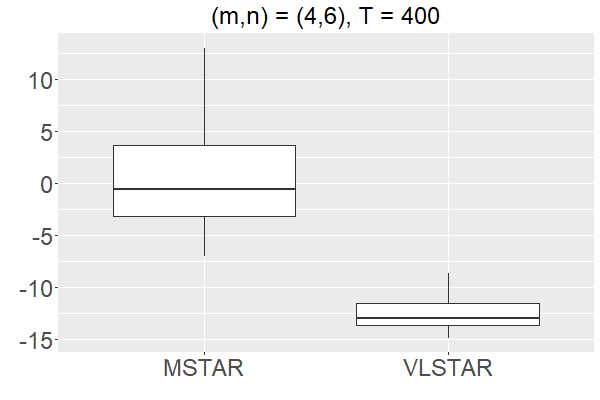} &\includegraphics[scale=0.25]{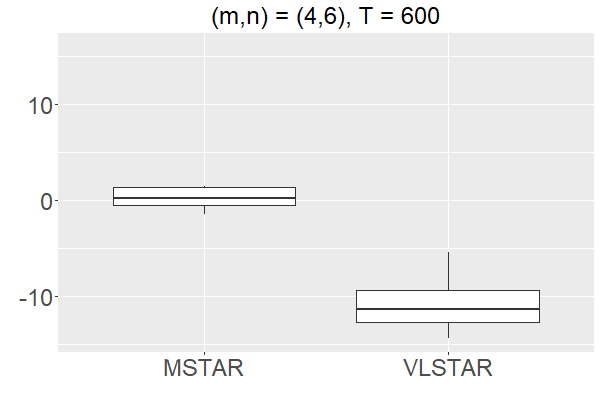} &\includegraphics[scale=0.25]{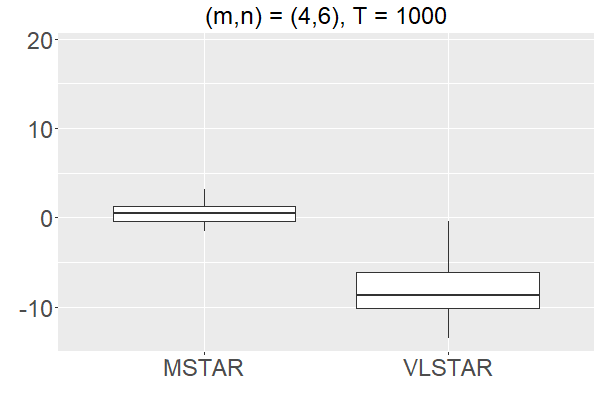}\\
	\end{tabular}
\end{figure}

Figure \ref{fig:Coefficients} shows the simulation results in terms of the measures in \eqref{eq:measure1} and \eqref{eq:measure2} for an increasing number of observations (from left to right) and an increasing dimension (from top to bottom) taking values in $(m,n) = (2,3)$, and $(4,6)$. Moreover, for each simulation setting, we report in Figure \ref{fig:Thresholds} and \ref{fig:Gammas} the box-plot of the difference of the estimated and observed threshold and slope parameter respectively in the 100 replications. 

From Figure \ref{fig:Coefficients} it emerges that, for each simulation setting, the MSTAR model strongly outperforms the VLSTAR model in the estimation of the coefficient matrices for both the regimes when the slope parameter is equal to 20 and the dimension of the matrix is $(4,6$). Moreover, it always overperforms the VLSTAR model in the estimation of the coefficient matrix in the first regime with mixed evidence in terms of the coefficient matrix in the second regime when data are simulated with $\gamma = 10$. 

The results in terms of estimated threshold in Figure \ref{fig:Thresholds} are similar. In this case, it can be observed that for $(m,n) = (4,6)$, the median of the difference between the estimated and the observed threshold is extremely close to zero regardless of the sample size (last row in the Figure), meaning that the threshold is correctly estimated. Overall, the estimated thresholds are in line with the observed value for both the methods, and the difference between the two methods is often negligible. Still, for larger matrices and larger samples, the MSTAR model seems to outperform the VLSTAR one also in terms of threshold estimation.

Figure \ref{fig:Gammas} reports the box-plots among the $R$ replicates of the slope parameter $\gamma$. It can be observed that the two models perform similarly well in the estimation of the slope parameter when $\gamma = 20$ and the dimension of the matrix is $(2,3)$. When the matrix dimension and the sample size increase, the difference between the estimated and the observed slope parameter is strictly close to zero for the MSTAR model, while it diverges for the VLSTAR.

Finally, Table \ref{table:MSE} highlights some interesting results. On the one hand, the MSE for the threshold parameter $c$ is systematically lower for the MSTAR model with respect to the results from the VLSTAR. On the other hand, with low-dimensional matrices (\textit{i.e.}, when $m = 2$ and $n = 3$) the latter exhibits a lower MSE when $\theta$ is equal to the slope parameter. Nevertheless, the reversed result is observed when the size of the $\mathbf{Y}$ matrix increases. 

\clearpage

\begin{table}[h]
	\centering
	\caption{\textbf{MSE over the $R = 100$ replicates for the estimated $\gamma$ and $c$.}}\label{table:MSE}
	\begin{tabularx}{0.75\textwidth}{lYYYYYY}
		\toprule
		& \multicolumn{6}{c}{Panel A: Slope parameter $\gamma = 10$ and dimension $(2,3)$}\\	
		& \multicolumn{2}{c}{$T = 400$} & \multicolumn{2}{c}{$T = 600$} & \multicolumn{2}{c}{$T = 1000$}\\
		\cmidrule{2-7}
		Model &    $c$     &  $\gamma$    &    $c$     &  $\gamma$  &    $c$     &  $\gamma$ \\
		\midrule
		MSTAR &  0.0027    &  21.5643     &  0.0015    &   16.5350  &   0.0010   &  7.4203\\
		VLSTAR&  0.0062    &  16.2638     &  0.0037    &   16.2595  &   0.0023   &  7.1958\\
			\toprule
	& \multicolumn{6}{c}{Panel B: Slope parameter $\gamma = 10$ and dimension $(4,6)$}\\	
	& \multicolumn{2}{c}{$T = 400$} & \multicolumn{2}{c}{$T = 600$} & \multicolumn{2}{c}{$T = 1000$}\\
			\cmidrule{2-7}
	Model &    $c$     &  $\gamma$    &    $c$     &  $\gamma$  &    $c$     &  $\gamma$ \\
	\midrule
	MSTAR &  0.0011    &  5.6837      &  0.0007    &   3.2501   &   0.0004   &  2.1458\\
	VLSTAR&  0.0122    &  14.0856     &  0.0091    &   12.7824  &   0.0059   &  10.7758\\
			\toprule
	& \multicolumn{6}{c}{Panel C: Slope parameter $\gamma = 20$ and dimension $(2,3)$}\\	
	& \multicolumn{2}{c}{$T = 400$} & \multicolumn{2}{c}{$T = 600$} & \multicolumn{2}{c}{$T = 1000$}\\
			\cmidrule{2-7}
	Model &    $c$     &  $\gamma$    &    $c$     &  $\gamma$  &    $c$     &  $\gamma$ \\
	\midrule
	MSTAR &  0.0007    &  123.7312    &  0.0004    &   101.7804 &   0.0002   &  57.2464\\
	VLSTAR&  0.0021    &  55.1497     &  0.0010    &   47.6796  &   0.0004   &  26.1468\\	
			\toprule
& \multicolumn{6}{c}{Panel D: Slope parameter $\gamma = 20$ and dimension $(4,6)$}\\	
& \multicolumn{2}{c}{$T = 400$} & \multicolumn{2}{c}{$T = 600$} & \multicolumn{2}{c}{$T = 1000$}\\
		\cmidrule{2-7}
Model &    $c$     &  $\gamma$    &    $c$     &  $\gamma$  &    $c$     &  $\gamma$ \\
\midrule
MSTAR &  0.0003    &  22.4654     &  0.0002    &   18.1687  &   0.0003   &  24.2003\\
VLSTAR&  0.0076    &  153.1464    &  0.0040    &   121.8717 &   0.0016   &  76.0079\\
		\bottomrule  
		\multicolumn{7}{p{0.7\textwidth}}{Note: Each entry in the Table reports the MSE computed between the parameter used for the simulation and the estimated parameter.}  
	\end{tabularx}
\end{table}

\section{Empirical applications}\label{sec:empirical}

\subsection{Economic indicators}

The data used for the first empirical application comprehends the monthly observations of the Economic Policy Uncertainty (EPU) index (first order difference) proposed by \cite{bake16}, the short term interest rate (first order log-difference) and the 10-Year interest rate (first order log-difference) from two countries, the United Kingdom and the United States, for the period that goes from January 2006 to September 2022. The data on EPU was retrieved from the `Economic Policy Uncertainty' site (\url{https://www.policyuncertainty.com/}), while the interest rates were collected from FRED. The series have been standardized before estimating the model. We choose to use the normalized temporal trend, $s_t = t/T$, as a transition variable for the MSTAR to identify a possible common structural break among the six time series considered. Furthermore, we use a diagonal matrix with $0.2$ entries for the initial values of $\mathbf{B}$, while the initial matrices in the second regime are diagonal matrices with $0.5$ on their diagonals. Following \cite{Chen2021}, the matrices $\mathbf{A}$ and $\mathbf{C}$ have been re-scaled to have a Frobenius norm equal to one and avoid lack of identification. The initial value for $\gamma$ is set equal to $5$, while the initial threshold is set to $0.75$.

\begin{figure}[h]
	\centering
	\caption{Plot of the time series used for the empirical analysis. In row, we report the economic indicators, while in column we report the country. We also report the estimated threshold for the transition variable (in dashed red).} 
	\label{fig:EmpiricalThreshold}
		\includegraphics[scale=0.55]{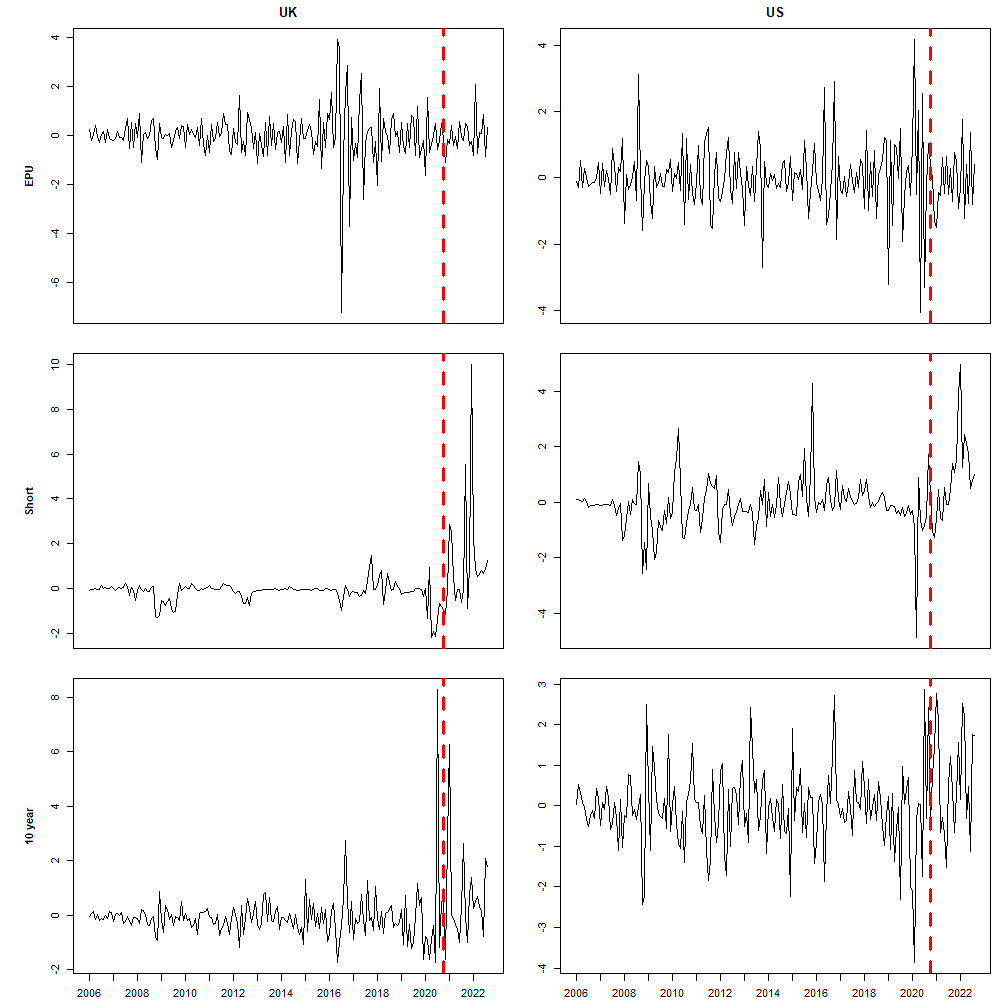}
\end{figure}

Figure \ref{fig:EmpiricalThreshold} reports the plot of the time series with the estimated threshold parameter for the transition variable. In this analysis, $\hat{c} = 0.906$, which means that the change of regime in the matrix-variate time series happens at November 2020. This is not surprising, since with the COVID-19 pandemic spread at the beginning of the 2020, investors moved to safer investments \citep{Janus2021}, and the uncertainty about economic policy exploded with the worsening of the economic status of countries \citep{Ahmed2021}. Moreover, this is almost the same estimate of the threshold (equal to $0.912$) obtained when a VLSTAR model is applied on the stacked vector of dependent variables, or when a matrix-variate threshold autoregressive model is applied ($\hat{c} = 0.897$).

The results in Tables \ref{table:MatrixA} and \ref{table:MatrixB} show that the cross-relations between the economic indicators and between countries are reinforced in the second regime. In fact, while the autoregressive coefficients remain mostly similar, the cross-effect becomes positive and significantly different from zero at any confidence level. If we analyse the single coefficient matrix, we observe that the only parameters that are significant in matrix $\mathbf{A}$ are those related to the effect of the lagged EPU and the interest rates on themselves. The scenario changes with the regime switch, since the EPU (first column) influences positively and significantly the growth of the interest rates at both short and long terms. This means that an higher economic policy uncertainty, in this case related to the COVID-19 pandemic, increases the interest rates growth. Moreover, the long-term interest rates growth is significantly associated with both the other economic indicators, negatively with the EPU and positively with the short-term interest rate growth.

When looking at the estimated matrices $\mathbf{B}$ and $\mathbf{D}$ in Table \ref{table:MatrixB}, it can be observed that the influence of the lagged economic indicators of UK on the indicators of US is positive for both the regimes but significant only for the second regime, and the same is true for the opposite direction. Overall, this example highlights the usefulness of a MSTAR model is twofold: on the one hand, it allows the estimation of a common structural break within the considered matrix-variate time series; on the other hand, several relations change with the regime-switching, meaning that a linear model would have inevitably wrongly under- or over-estimate these relations.

\clearpage

\begin{table}[h]
	\centering
	\caption{Estimated coefficient matrices $\mathbf{A}$ and $\mathbf{C}$ for both the regimes. $p$-values are reported between brackets.}\label{table:MatrixA}
	\begin{tabular}{lrrr@{\hskip 0.8cm}rrr}
		\toprule
		& \multicolumn{3}{c}{Regime 1 (matrix $\mathbf{A}$)}          & \multicolumn{3}{c}{Regime 2 (matrix $\mathbf{C}$)}\\
		& EPU   &  Short-term  & 10-year        & EPU   &  Short-term  & 10-year\\
		\midrule
		EPU        &  $\underset{(3.66e-16)}{-0.524}$ & $\underset{(0.564)}{0.040}$    & $\underset{(0.184)}{-0.082}$    & $\underset{(1.01e-12)}{-0.374}$ & $\underset{(0.003)}{0.136}$    & $\underset{(5.63e-09)}{-0.260}$  \\  
		Short-term &  $\underset{(0.181)}{-0.094}$    & $\underset{(3.25e-22)}{0.636}$ & $\underset{(3.06e-11)}{0.479}$  & $\underset{(9.16e-11)}{0.384}$  & $\underset{(4.37e-11)}{0.321}$ & $\underset{(0.798)}{0.010}$\\
		10-year    &  $\underset{(0.101)}{-0.101}$    & $\underset{(0.006)}{-0.194}$   & $\underset{(0.007)}{0.166}$     & $\underset{(1.19e-18)}{0.611}$  & $\underset{(0.276)}{0.045}$    & $\underset{(4.56e-16)}{0.385}$\\		
		\bottomrule
	\end{tabular}
\end{table}

\begin{table}[h]
	\centering
	\caption{Estimated coefficient matrices $\mathbf{B}$ and $\mathbf{D}$ for both the regimes. $p$-values are reported between brackets.}\label{table:MatrixB}
	\begin{tabular}{lrr@{\hskip 0.3cm}rr}
		\toprule
		& \multicolumn{2}{c}{Regime 1 (matrix $\mathbf{C}$)}   & \multicolumn{2}{c}{Regime 2 (matrix $\mathbf{D}$)}\\
		& UK        & US       & UK        & US\\
		\midrule
		UK      & $\underset{(8.51e-17)}{0.536}$ & $\underset{(0.927)}{-0.037}$    & $\underset{(8.45e-23)}{-0.840}$ & $\underset{(0.084)}{0.754}$\\
		US      & $\underset{(0.856)}{0.071}$    & $\underset{(2.88e-11)}{0.427}$  & $\underset{(0.449)}{0.615}$ & $\underset{(2.69e-27)}{0.881}$ \\		
		\bottomrule
	\end{tabular}
\end{table}

\subsection{Air pollution levels in London}
We further apply the proposed model to the daily levels of four pollutants, the nitrogen dioxide (NO$_2$), the trioxygen (or ozone, O$_3$), the suspended particulate matter with a diameter of 10 micrometers or less (PM$_{10}$), and the sulfur dioxide (SO$_2$), in the period from January 1, 2019 to December 31, 2019 from three air quality stations in the city of London. The data were collected from the website of the European Environment Agency (EEA). The series have been standardized before estimating the model. As candidate transition variables, we selected the lagged average temperature and the lagged precipitation level at Heathrow airport.  Temperature and precipitation data was retrieved from the National Centers for Environmental Information of the National Oceanic and Atmospheric Administration (NOAA).

We are supposing that the pollution levels from the different monitoring stations in the city of London move between two regimes \citep{Unal2011}. As before, we use a diagonal matrix with $0.2$ entries for the initial values of $\mathbf{B}$, while the initial matrices in the second regime are diagonal matrices with $0.5$ on their diagonals. Following \cite{Chen2021}, the matrices $\mathbf{A}$ and $\mathbf{C}$ have been re-scaled to have a Frobenius norm equal to one and avoid lack of identification.
The initial value for $\gamma$ is set equal to $5$, while the initial threshold is equal to the third quartile of the distribution of $s_t$.

\begin{figure}[h]
	\centering
	\caption{Plot of the time series used for the empirical analysis on air pollution data when the transition variable is the lagged average temperature. In row, we report the pollutant, while in column we report the monitoring station. We also report the estimated vertical lines identifying the second regime (in dashed grey).} 
	\label{fig:EmpiricalPoll_temp}
	\includegraphics[scale=0.55]{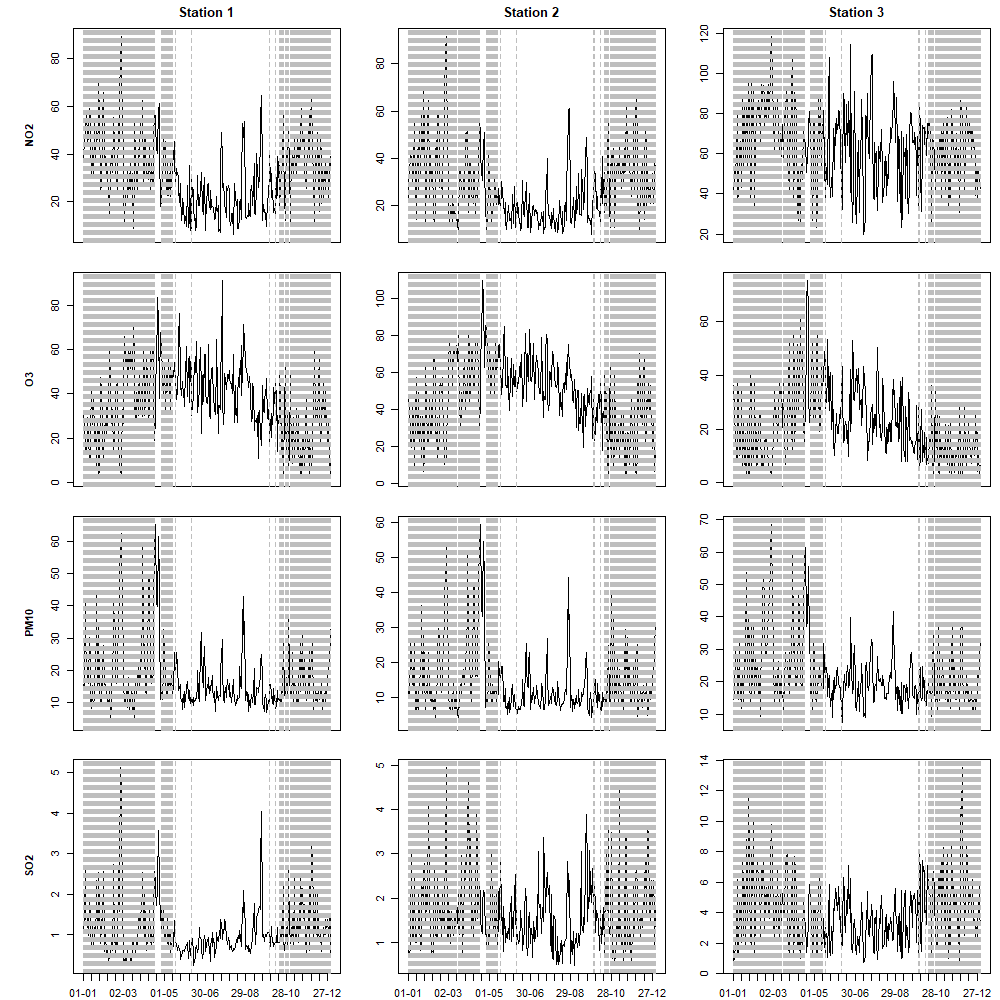}
\end{figure}

Figure \ref{fig:EmpiricalPoll_temp} reports the observed time series with the regimes detected through the lagged average temperature as a transition variable, while the estimated parameters are reported in Tables \ref{table:MatrixA2} and \ref{table:MatrixB2}. We observe that the parameters in $\mathbf{A}$ are not statistically significant, while the autoregressive parameters in $\mathbf{C}$ (\textit{i.e.}, the values on the diagonal) are mostly significant. As already seen in the previous application, there emerges a different behavior of the dependencies between the two regimes. In the second regime, a cross-effect between NO$_2$, PM$_{10}$, and SO$_{2}$ seems to be present. Similarly, the autoregressive effects between stations become significant only in the second regime.

Different results are obtained when the lagged precipitation level is used as a transition variable in Figure \ref{fig:EmpiricalPoll_temp}, and Tables \ref{table:MatrixA3} and \ref{table:MatrixB3}. In this case, the autoregressive coefficients of the pollutants are significant also in the first regime, while the coefficients of the monitoring stations are close to the nonstationary level in the second regime.

\begin{table}[h]
	\centering
	\caption{Estimated coefficient matrices $\mathbf{A}$ and $\mathbf{C}$ for both the regimes in the empirical analysis on air pollution data when the transition variable is the lagged average temperature. $p$-values are reported between brackets.}\label{table:MatrixA2}
	\begin{tabular}{lrrrr@{\hskip 0.8cm}rrrr}
		\toprule
		& \multicolumn{4}{c}{Regime 1 (matrix $\mathbf{A}$)}          & \multicolumn{4}{c}{Regime 2 (matrix $\mathbf{C}$)}\\
		&NO$_2$   &  O$_3$  & PM$_{10}$ & SO$_2$      &NO$_2$   &  O$_3$  & PM$_{10}$ & SO$_2$\\
		\midrule
		NO$_2$        &  $\underset{(0.311)}{0.532}$ & $\underset{(0.890)}{0.072}$ & $\underset{(0.948)}{-0.034}$ & $\underset{(0.972)}{-0.019}$ & $\underset{(2.13e-16)}{-0.277}$ & $\underset{(0.641)}{0.014}$    & $\underset{(9.28e-10)}{0.195}$ & $\underset{(8.57e-24)}{0.362}$  \\  
		O$_3$         &  $\underset{(0.967)}{0.022}$ & $\underset{(0.307)}{0.552}$ & $\underset{(0.844)}{0.106}$  & $\underset{(0.992)}{-0.006}$ & $\underset{(0.062)}{0.051}$    & $\underset{(9.81e-31)}{-0.405}$ & $\underset{(0.101)}{0.045}$    & $\underset{(7.41e-50)}{-0.630}$\\
		PM$_{10}$     &  $\underset{(0.763)}{0.158}$ & $\underset{(0.857)}{0.095}$ & $\underset{(0.349)}{0.491}$  & $\underset{(0.917)}{-0.055}$ & $\underset{(1.19e-10)}{-0.206}$ & $\underset{(0.122)}{-0.046}$    & $\underset{(0.529)}{0.019}$    & $\underset{(1.62e-11)}{0.217}$\\
		SO$_2$        &  $\underset{(0.888)}{0.076}$ & $\underset{(0.982)}{-0.012}$& $\underset{(0.954)}{-0.031}$ & $\underset{(0.541)}{0.330}$  & $\underset{(0.005)}{-0.078}$    & $\underset{(0.771)}{0.008}$    & $\underset{(0.157)}{0.039}$    & $\underset{(1.68e-20)}{0.297}$\\		
		\bottomrule
	\end{tabular}
\end{table}

\begin{table}[h]
	\centering
	\caption{Estimated coefficient matrices $\mathbf{B}$ and $\mathbf{D}$ for both the regimes in the empirical analysis on air pollution data when the transition variable is the lagged average temperature. $p$-values are reported between brackets.}\label{table:MatrixB2}
	\begin{tabular}{lrrr@{\hskip 0.6cm}rrr}
		\toprule
		& \multicolumn{3}{c}{Regime 1 (matrix $\mathbf{C}$)}   & \multicolumn{3}{c}{Regime 2 (matrix $\mathbf{D}$)}\\
		& Station 1        & Station 2   & Station 3       & Station 1        &  Station 2  & Station 3\\
		\midrule
		Station 1      & $\underset{(0.667)}{0.226}$ & $\underset{(0.983)}{0.012}$ & $\underset{(0.960)}{-0.032}$ & $\underset{(3.06e-16)}{0.274}$ & $\underset{(0.795)}{0.030}$ & $\underset{(9.36e-04)}{0.407}$\\
		Station 2      & $\underset{(0.917)}{0.053}$ & $\underset{(0.703)}{0.176}$ & $\underset{(0.963)}{-0.030}$ & $\underset{(0.110)}{0.239}$    & $\underset{(0.026)}{0.065}$ & $\underset{(0.003)}{0.365}$ \\
		Station 3      & $\underset{(0.928)}{0.044}$ & $\underset{(0.962)}{0.023}$ & $\underset{(0.831)}{0.135}$  & $\underset{(0.072)}{0.266}$    & $\underset{(0.228)}{0.127}$ & $\underset{(7.54e-33)}{0.513}$ \\		
		\bottomrule
	\end{tabular}
\end{table}

\begin{figure}[h]
	\centering
	\caption{Plot of the time series used for the empirical analysis on air pollution data when the transition variable is the lagged precipitation level. In row, we report the pollutant, while in column we report the monitoring station. We also report the estimated vertical lines identifying the second regime (in dashed grey).} 
	\label{fig:EmpiricalPoll_prec}
	\includegraphics[scale=0.55]{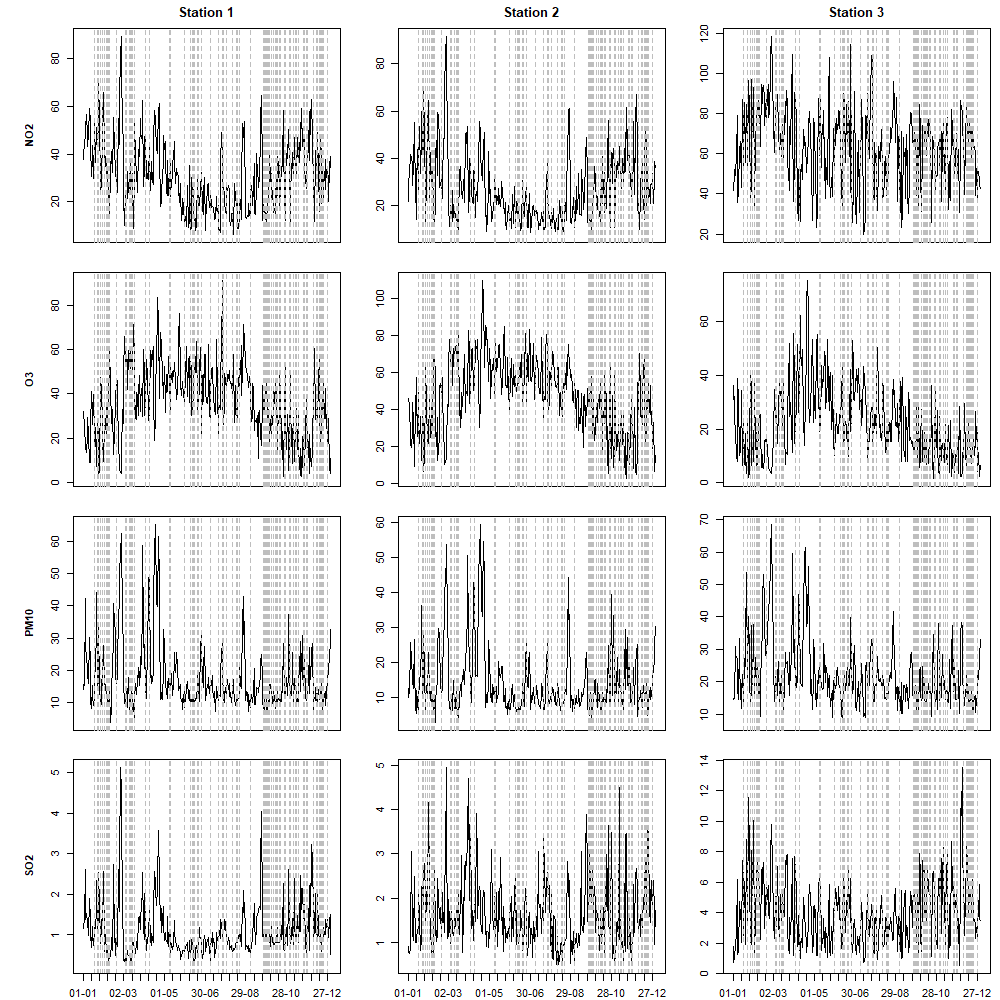}
\end{figure}

\begin{table}[h]
	\centering
	\caption{Estimated coefficient matrices $\mathbf{A}$ and $\mathbf{C}$ for both the regimes in the empirical analysis on air pollution data when the transition variable is the lagged precipitation level. $p$-values are reported between brackets.}\label{table:MatrixA3}
	\begin{tabular}{lrrrr@{\hskip 0.8cm}rrrr}
		\toprule
		& \multicolumn{4}{c}{Regime 1 (matrix $\mathbf{A}$)}          & \multicolumn{4}{c}{Regime 2 (matrix $\mathbf{C}$)}\\
		&NO$_2$   &  O$_3$  & PM$_{10}$ & SO$_2$      &NO$_2$   &  O$_3$  & PM$_{10}$ & SO$_2$\\
		\midrule
NO$_2$   &  $\underset{(0.004)}{0.442}$ & $\underset{(0.451)}{0.093}$    & $\underset{(0.720)}{0.044}$  & $\underset{(0.608)}{-0.063}$    & $\underset{(4.37e-23)}{0.116}$ & $\underset{(7.33e-06)}{0.046}$ & $\underset{(2.48e-19)}{-0.102}$ & $\underset{(4.40e-33)}{0.543}$  \\  
O$_3$    &  $\underset{(0.812)}{0.025}$ & $\underset{(7.45e-05)}{0.422}$ & $\underset{(0.233)}{0.126}$  & $\underset{(0.059)}{-0.201}$    & $\underset{(0.040)}{-0.021}$    & $\underset{(0.302)}{0.011}$    & $\underset{(3.51e-05)}{0.044}$ & $\underset{(7.29e-26)}{-0.799}$\\
PM$_{10}$&  $\underset{(0.389)}{0.106}$ & $\underset{(0.614)}{0.062}$    & $\underset{(0.014)}{0.305}$  & $\underset{(2.64e-06)}{-0.584}$ & $\underset{(0.055)}{0.019}$    & $\underset{(0.763)}{-0.003}$    & $\underset{(2.84e-04)}{0.037}$ & $\underset{(1.75e-17)}{0.169}$\\
SO$_2$   &  $\underset{(0.919)}{0.011}$ & $\underset{(0.974)}{0.004}$    & $\underset{(0.957)}{-0.006}$ & $\underset{(0.002)}{0.325}$     & $\underset{(0.639)}{0.005}$    & $\underset{(0.864)}{0.002}$    & $\underset{(0.612)}{-0.005}$    & $\underset{(7.56e-14)}{0.084}$\\		
		\bottomrule
	\end{tabular}
\end{table}

\begin{table}[h]
	\centering
	\caption{Estimated coefficient matrices $\mathbf{B}$ and $\mathbf{D}$ for both the regimes in the empirical analysis on air pollution data when the transition variable is the lagged precipitation level. $p$-values are reported between brackets.}\label{table:MatrixB3}
	\begin{tabular}{lrrr@{\hskip 0.6cm}rrr}
		\toprule
		& \multicolumn{3}{c}{Regime 1 (matrix $\mathbf{C}$)}   & \multicolumn{3}{c}{Regime 2 (matrix $\mathbf{D}$)}\\
		& Station 1        & Station 2   & Station 3       & Station 1        &  Station 2  & Station 3\\
		\midrule
		Station 1      & $\underset{(0.755)}{0.038}$  & $\underset{(0.666)}{0.132}$ & $\underset{(0.980)}{0.019}$ & $\underset{(2.14e-77)}{0.980}$ & $\underset{(0.315)}{-0.334}$    & $\underset{(0.674)}{-0.259}$\\
		Station 2      & $\underset{(0.542)}{-0.143}$ & $\underset{(0.008)}{0.315}$ & $\underset{(0.960)}{0.032}$ & $\underset{(6.68e-08)}{0.949}$ & $\underset{(6.39e-95)}{-0.489}$ & $\underset{(0.434)}{-0.407}$ \\
		Station 3      & $\underset{(0.496)}{-0.121}$ & $\underset{(0.739)}{0.073}$ & $\underset{(0.096)}{0.218}$ & $\underset{(9.12e-11)}{0.917}$ & $\underset{(0.216)}{0.253}$     & $\underset{(4.30e-82)}{-0.393}$ \\		
		\bottomrule
	\end{tabular}
\end{table}
\clearpage

\section{Conclusions}\label{sec:Conclusions}
In this article, we have proposed a smooth transition autoregressive model for matrix-variate time series which extends the multivariate model from \cite{teya14}. In a numerical study, we show that the MSTAR model is able to outperform the stacked vectorization form through a VLSTAR in terms of estimated coefficients, threshold and slope parameters. This is mostly true for larger samples and larger matrices. We have also applied the newly proposed MSTAR model to economic indicators from the United Kingdom and the United States, showing that the model is also capable of detecting common structural breaks determining the regime switches. We further used the MSTAR model with air pollution data using a stationary transition variable.

Since the literature on matrix-variate problems is new, there are several open questions that could be solved in future works. For instance, a linearity test of the matrix-variate time series could avoid the use of a more complex specification when a linear one is needed. Moreover, the choice of the number of regimes is crucial in this kind of literature \citep{Liu2022} and a proper procedure for the detection of the regimes in a smooth transition model would be helpful.

\section*{Acknowledgments}
The author acknowledges that the research is part of the project PON "Ricerca e Innovazione" 2014-2020 - Azione I.2 "Mobilità dei Ricercatori" - Avviso di cui al D.D. n. 407 del 27 febbraio 2018 e D.D. n. 1621 del 12/08/2019 "Attraction and International Mobility" - LINEA 1 (Mobilità dei Ricercatori) (CUP n. D24I19001680001 - Id. proposta: AIM1894803 Linea 1: attività 3 Area di specializzazione SNSI: Smart, Secure and Inclusive Communities).

%
%
%
%
%

\bibliographystyle{Chicago}

\bibliography{Econometrics}

\end{document}